\documentclass[preprint,a4paper,12pt]{elsarticle}

\usepackage{amsmath}
\usepackage{graphicx}
\usepackage{dcolumn}
\usepackage{bm}
\usepackage{caption}
\usepackage{float}
\usepackage{dsfont}
\usepackage{mathrsfs}
 \RequirePackage{mathptmx} 
\usepackage[T1]{fontenc}
\usepackage{grfext}
\usepackage{multirow}
\usepackage{makecell}
\usepackage{threeparttable}
\usepackage{float}\usepackage{xcolor}
\usepackage{booktabs}\usepackage{multicol}
\usepackage{ragged2e}
\usepackage[bookmarks=true,hidelinks,citecolor=blue,linkcolor=blue,colorlinks=true]{hyperref}

\usepackage[margin=0.9in]{geometry}\usepackage{subcaption}
\usepackage{subcaption}
\usepackage[T1]{fontenc}
\usepackage{subcaption}
\journal{Chinese journal of Physics}

\begin{document}

\begin{frontmatter}

\title{Investigation of Nuclear Structure and $\beta$-decay Properties of As Isotopes }
\author{Jameel-Un Nabi $^{1}$, Abdul Kabir $^{2}$, Wajeeha Khalid $^{1}$, Syeda Anmol Rida $^{1}$, Izzah Anwaar$^{2}$}
\address{Corresponding Author Email; wajeeha.khalid@uow.edu.pk}
\address{$^{1}$University of Wah, Quaid Avenue, Wah Cantt 47040, Punjab, Pakistan.}
\address{$^{2}$ Department of Space Science, Institute of Space Technology, Islamabad 44000, Pakistan.}

\begin{abstract}
The nuclear ground state properties of $^{67-80}$As nuclei have been investigated within the framework of relativistic mean field (RMF) approach.  The RMF model with density-dependent  (DD-ME2) interaction is utilized for the calculation of  potential energy curves and the nuclear ground-state deformation parameters ($\beta_2$) of selected As isotopes. Later, the $\beta$-decay properties of As isotopes were studied using the proton-neutron quasi particle random phase approximation (pn-QRPA) model. These include Gamow Tellar (GT) strength distributions, log $\textit{ft}$ values, $\beta$-decay half-lives, stellar $\beta^\pm$ decays and stellar electron/positron capture rates. The $\beta_{2}$ values computed from RMF model were employed in the pn-QRPA model as an input parameter for the calculations of $\beta$-decay properties for $^{67-80}$As. The calculated log~$\textit{ft}$ values were in decent agreement with the measured data.  The predicted $\beta$-decay half-lives matched the experimental values within a factor of 10. The stellar rates were compared with the shell model results. Only at high temperature and density values, the sum of $\beta^+$ and electron capture rates had a finite contribution. On the other hand, the sum of $\beta^-$ and positron capture rates were sizeable only at low density and high temperature values.  For all such cases, the pn-QRPA rates were found to be bigger than the shell model rates up to a factor of 33 or more.  The findings reported in the current investigation could prove valuable for simulating the late-stage stellar evolution of massive stars.
\end{abstract}
\begin{keyword}
pn-QRPA, $\beta$-decay properties, GT strength distribution, Deformation parameter, RMF model, log $\textit{ft}$, Stellar rates.
\end{keyword}
\end{frontmatter}
\newpage
\section{Introduction}
	\label{intro}
Accurate measurements and reliable theoretical calculations of low-energy nuclear structure properties play a vital role in the modeling and comprehension of fundamental nuclear phenomena. These include $\beta$-decays which are closely linked to  the stellar nucleosynthesis. The unique features of low-lying excitation energies and decay patterns found in neutron-rich nuclei around $A = 100$ and possessing $N\approx$ 60  are especially intriguing, holding importance for both nuclear structure studies and astrophysical investigations~\cite{Nomu22}. In these nuclei, both shape transitions and shape coexistence may occur. Additionally, these characteristics contribute to the rapid neutron capture ($r$-) process, accounting for the creation of massive elements through nucleosynthesis in explosive environments. The $r$-process involves multiple neutron captures on iron until $\beta$-decay leads to a change in atomic number, thereby facilitating further neutron captures on the new specie. The $r$-process is accountable for creating about fifty percent of the nuclei that are heavier than iron~\cite{Bur57, Woo94}. The exact astrophysical origin of the $r$-process  remains uncertain.
The radioactive ion beams, at major experimental facilities worldwide, have been extensively utilized for precise measurement of $\beta$-decay half-lives of neutron-rich nuclei. Physicists have  investigated the $\beta$-decay characteristics of various neutron-rich isotopes of Germanium (Ge) and Arsenic (As)~\cite{Maz13, Hof81}. Most of the $\beta$-decay rates of neutron-rich nuclei have been investigated through theoretical models, primarily because of the limited availability of experimental data. Theoretical efforts to describe both low-lying nuclear states and their decay characteristics  represent a significant challenge. The decay process has been examined theoretically using various nuclear models. Few commonly used models to perform these calculations would include the interacting boson model (IBM)~\cite{Iac80}, the quasiparticle random phase approximation (QRPA) ~\cite{Aly04, Min21}, and the large-scale shell model (LSSM)~\cite{Kum20}. The structure of odd-A energy levels and the electromagnetic properties of $^{69, 71, 73}$As and $^{69, 71, 73}$Ge were investigated in the framework of the proton-neutron interacting boson-fermion model~\cite{Bra04}. Furthermore, the $\beta$-decay properties of neutron-rich isotopes were explored using a deformed proton-neutron quasi-particle random-phase approximation~\cite{Sar15}. Recently, the $\beta$-decay and electron capture rates of Manganese (Mn) isotopes in the mass range  $A=53-63$ were calculated using the pn-QRPA approach~\cite{She23}. On the other hand, nuclear structure properties and $\beta$-decay half-lives of exotic proton-rich waiting point nuclei in the mass region $A\approx$ 70 were investigated using both the proton-neutron quasiparticle random phase approximation (pn-QRPA) and the interacting boson model-1 (IBM-1)~\cite{Nab16}.

We utilize both the relativistic mean field (RMF)~\cite{Wal74} and pn-QRPA models to provide a reliable description of the nuclear structure and $\beta$-decay properties of As isotopes.
The pn-QRPA theory is capable of performing a microscopic calculation of weak rates for any arbitrary heavy nucleus. Klapdor et al.~\cite{Kla84} were the first to conduct calculation of $\beta$-decay rates for many nuclei far-off from the line of stability using a microscopic nuclear theory. This model was later utilized to calculate the $\beta$-decay half-lives of $\sim$ 6000 neutron-rich nuclei, covering a broad range from the neutron drip line to the line of stability~\cite{Sta90}. The same pn-QRPA model, with a separable and schematic interaction, was used to perform a microscopic calculation of  $\beta^{+}$/electron capture rates of neutron-deficient nuclei with atomic numbers ranging from $Z$\,=\,10~-~108, extending up to the proton drip line for over 2000 nuclei ~\cite{Hir93}. The weak interaction rates have contributed significantly to our understanding of the $r$-process. Fuller et al.~\cite{Ful82} conducted a thorough effort to compile these rates at stellar temperatures and densities, with a specific emphasis on decays from excited states of parent nuclei.

In the present work, we calculate ground-state properties  of selected As isotopes, including binding energies and nuclear deformation parameter ($\beta_{2}$),  within the RMF framework \cite{Walecka} with  DD-ME2 interaction~\cite{Lal05}. The RMF framework describes various ground-state nuclear properties of nuclei using a phenomenological approach. For further description of the nuclear model, one may see Ref.~\cite{Lal05}. The $\beta$-decay properties of  As nuclei, in the region $67 \leq A \leq 80$, have been studied recently in the framework of the particle-core coupling scheme and nuclear density functional theory~\cite{Nom22}. We  investigate the $\beta$-decay properties of selected As isotopes (including the terrestrial $\beta$-decay half-lives, GT strength distributions, log $\textit{ft}$ values and stellar rates) by utilizing the theoretical framework of the deformed pn-QRPA model. The present model-based analysis is compared with previously observed and predicted data.
 
The paper is organized as follows. In Section 2, we provide a brief explanation of the RMF and pn-QRPA models used to calculate the nuclear structure and $\beta$-decay properties, respectively. Section 3 presents our results with relevant discussion.  Section 4 concludes the findings of the current investigation.

\section{Formalism}
 
\subsection{\textbf{The RMF framework}}
In the RMF model, nucleons interact by exchanging mesons and photons~\cite{Walecka}. The preliminary RMF model struggled to describe nuclear surface features and the incompressibility of nuclear matter. For this reason, a nonlinear form of the model was developed (see Ref. \cite{rin96} and references therein). The later versions of the model were termed  as covariant density functional theory and included density-dependent meson-exchange and density-dependent point-coupling models \cite{Nikšić}. In the present investigation, the ground state parameters for nuclei have been determined by employing the density-dependent meson-exchange (DD-ME2) \cite{Lal05} version of the RMF model.
Additional information regarding different RMF model versions may be found in \cite{men06}. 
The following phenomenological Lagrangian density was considered for solution
\begin{equation}
	\mathcal{L}=\mathcal{L}_N+\mathcal{L}_m+\mathcal{L}_{int},\label{lagden}
\end{equation}  
where $\mathcal{L}_N$ represents the field of free nucleons, defined below
\begin{equation}
	\mathcal{L}_N=\bar{\Psi}(i\gamma_\mu\partial^\mu-m)\Psi,\label{lagnuc}
\end{equation}
$\psi$ refers to the Dirac spinor, while $m$ stands for the nucleon's mass.  $\mathcal{L}_m$ represents the meson and electromagnetic fields.
\footnotesize
 \begin{eqnarray}
	\mathcal{L}_m &=\frac{1}{2}\partial_\mu\sigma\partial^\mu\sigma-\frac{1}{2}m_\sigma^2\sigma^2-\frac{1}{2}\Omega_{\mu\nu}\Omega^{\mu\nu}+\frac{1}{2}m^2_\omega\omega_\mu\omega^\mu \nonumber \\ 
	&-\frac{1}{4}\overrightarrow{R}_{\mu\nu}.\overrightarrow{R}^{\mu\nu}+\frac{1}{2}m_\rho^2 \overrightarrow{\rho}_\mu.\overrightarrow{\rho}^\mu -\frac{1}{4}F_{\mu\nu}F^{\mu\nu}, \label{lagmes}
\end{eqnarray}
\normalsize
 where arrows indicate isovectors. $m_\sigma$, $m_\omega$ and $m_\rho$ represent the masses of the related mesons while $\Omega_{\mu\nu}$, $\overrightarrow{R}_{\mu\nu}$ and $F_{\mu\nu}$ are field tensors. Similarly, $\mathcal{L}_{int}$ includes both the photon-nucleon  and meson-nucleon interactions. The explicit forms of the three terms are given below 
\footnotesize
\begin{eqnarray}
	\mathcal{L}_{int}&= -g_\sigma\bar{\Psi}\Psi \sigma - g_\omega\bar{\Psi}\gamma^\mu\Psi\omega_\mu - g_\rho\bar{\Psi}\overrightarrow{\tau}\gamma^\mu\Psi.\overrightarrow{\rho}_\mu \nonumber \\ 
	&-e\bar{\Psi}\gamma^\mu\Psi A_\mu.
	\label{lagrangian}
\end{eqnarray}
\normalsize
The coupling constants for the relevant mesons are represented by $g_\sigma$, $g_\omega$, and $g_\rho$.
Usage of the Lagrangian density yields the Hamiltonian density for the static case. For further details, one may see Ref.~\cite{rin96}.
The quadrupole moment constrained RMF calculation was performed to obtain the potential energy curves (PECs) as a function of $\beta_{2}$. For the PEC analysis, we employed the Bardeen-Cooper-Schrieffer (BCS) constant-gap approximation with empirical pairing gaps \cite{pm}.

Studying even-even systems within the mean field approach is a good approximation. In this case, the configurations, neglected above the mean field ground state, are 4- or higher-quasiparticle (qp) configurations. The 2-qp configurations do not couple to the Hamiltonian ($H_{20}=0$). Mixing configurations are relatively few and separated by the pairing gaps.  The exact solution has only a small admixture of higher qp configurations (4-qp and higher). On the other hand, investigating odd-A nuclei using the RMF model is rather challenging. In this case, there are many 3-qp states in the region close to the ground state which may mix. The pairing gap even increases the level density of neighboring 1-qp states. Only a few of the $H_{31}$ matrix elements vanish. The mean field approximation in odd-A cases is not as good as in even-even nuclei. Of course, sometimes symmetries help ($K$-value in deformed nuclei), but this is not always the case. We performed the HFB-calculations in odd-A systems by using the blocking technique. The blocking was carried out by replacing one $U$-vector with the corresponding $V$-vector~(see section~6.3.2 of Ref.~\cite{Rin04}). We carried out the blocking calculations with small modifications of the HFB-code \cite{HBF}.

\subsection{\textbf{The pn-QRPA Model}}
The pn-QRPA model was employed for the calculation of $\beta$-decay properties of the As nuclei. The Hamiltonian of the pn-QRPA model was described using 
\begin{equation} \label{eq1}
	\mathscr{H}^{QRPA} = \mathscr{H}^{sp}+\mathscr{V}^{pair }+\mathscr{V}_{GT} ^{pp}+\mathscr{V}_{GT}^{ph}.
\end{equation}
The Hamiltonian for a single particle is denoted as $\mathscr{H}^{s p}$, while $\mathscr{V}^{\text {pair }}$ represents the interaction between nucleons. The terms $\mathscr{V}_{G T}^{p p}$ and $\mathscr{V}_{G T}^{p h}$ correspond to interactions involving particle-particle ($pp$) and particle-hole ($ph$) GT interactions, respectively. Wave functions and energies of single particle were calculated via the Nilsson model \cite{Nil55}. The oscillator constant for nucleon was determined using $\hbar\omega=(45A^{-1/3}-25A^{-2/3})$. Other parameters of the model include $\beta_{2}$, the pairing gaps, parameters of the Nilsson potential (NPP), GT force parameters and the $Q$-values. The NPP were adopted from Ref.~\cite{Rag84} and $Q$-values were computed from the mass excess values ~\cite{Kon21}.   In order to obtain proton and neutron quasiparticle energies and occupation probabilities, the BCS equations were solved with pairing gaps computed using the relation
\begin{equation} \label{eq2}
	\Delta _{nn} = \frac{1}{8}(-1)^{A-Z+1}[2 S_{n}(A+1,Z)-4S_{n}(A,Z)+2S_{n}(A-1,Z)]
\end{equation}
\begin{equation} \label{eq3}
	\Delta _{pp} = \frac{1}{8}(-1)^{1+Z}[2 S_{p}(A+1,Z+1)-4S_{p}(A,Z)+2S_{p}(A-1,Z-1)],
\end{equation}
where $S_{p} (S_{n})$ is the separation energy of
protons (neutrons) in units of MeV.
As mentioned earlier, the $\beta_{2}$ values were determined using the RMF model. 
The GT force parameters were taken from Ref.~\cite{Hom96}. For details of the pairing force and residual interaction as well as solution of Eq.~\ref{eq1}, please see Ref.~\cite{Mut92}. 

The partial half-lives $t_{1/2}$ were calculated using the relation
\small
	\begin{eqnarray}\label{eq12}
		t_{1/2} = \frac{C}{(g_A/g_V)^2f_A(A, Z, E){B}_{GT}(\omega)+f_V(A, Z, E){{B}_F(\omega)}},
	\end{eqnarray}
	\normalsize
	where  C (= ${2\pi^3 \hbar^7 ln2}/{g^2_V m^5_ec^4}$) was taken as 6143 s \cite{Har09}. The ratio  $g_A/g_V$ was taken as -1.2694 \cite{Nak10} and $E$ = ($Q$ - $\omega$), where $Q$ represent the $Q$-value of the reaction. $\omega$ is the excitation energy of the QRPA phonon determined by solving the RPA matrix equation.  $f_V(A, Z, E)$  and $f_A(A, Z, E)$ are the phase space integrals   for vector transition and axial vector,  respectively. These were computed using the recipe given in Ref.~\cite{Gov71}. ${B}_{GT}$ (${B}_{F}$) are the reduced GT (Fermi) transition probabilities. 
	The terrestrial half-lives of $\beta$-decay were calculated by summing over the inverses of partial half-lives and taking inverse
	\begin{equation}\label{eq13}
		~ ~ ~ ~ ~ ~ ~ ~ ~ ~ ~ ~ ~	T_{1/2} = \left(\sum_{0 \le \omega \le Q} (\frac{1}{t_{1/2}})\right)^{-1}.
	\end{equation}
	
The stellar weak rate, from parent level $n$ to daughter state $m$,  was computed using
\footnotesize
	\begin{equation} \label{eq14}
		~ ~ ~ ~ ~ ~ ~ ~ ~ ~ ~ ~ ~	\lambda _{mn}^{\beta^{\pm} /EC/PC} =\ln 2\frac{f_{mn}^{\beta^{\pm} /EC/PC}(\rho,T
			,E_{f})}{(ft)_{mn}},
	\end{equation}
	\normalsize
where the numerator is the phase space integral for the weak decay reactions and the denominator is the $ft$ value of the transition related to the calculation of reduced transition probabilities of the charge-changing transitions. Details of the solution of Eq.~\ref{eq14} may be seen from Refs.~\cite{Nab99, Nab04}.

The total stellar rates were determined using
\begin{equation} \label{eq48}
		~ ~ ~ ~ ~ ~ ~ ~ ~ ~ ~ ~ ~	\lambda^{\beta^{\pm} /EC/PC} =\sum _{mn}P_{m} \lambda _{mn}^{\beta^{\pm} /EC/PC},
\end{equation}
where $P_m$ stands for the occupation probability of the parent excited state and was determined using the Boltzmann distribution. The summation over initial and final states was carried out until desired convergence level was achieved in our rate calculation.
	
\section{Results and Discussion}
\begin{figure}[H]
	\centering
	\includegraphics[width=13cm]{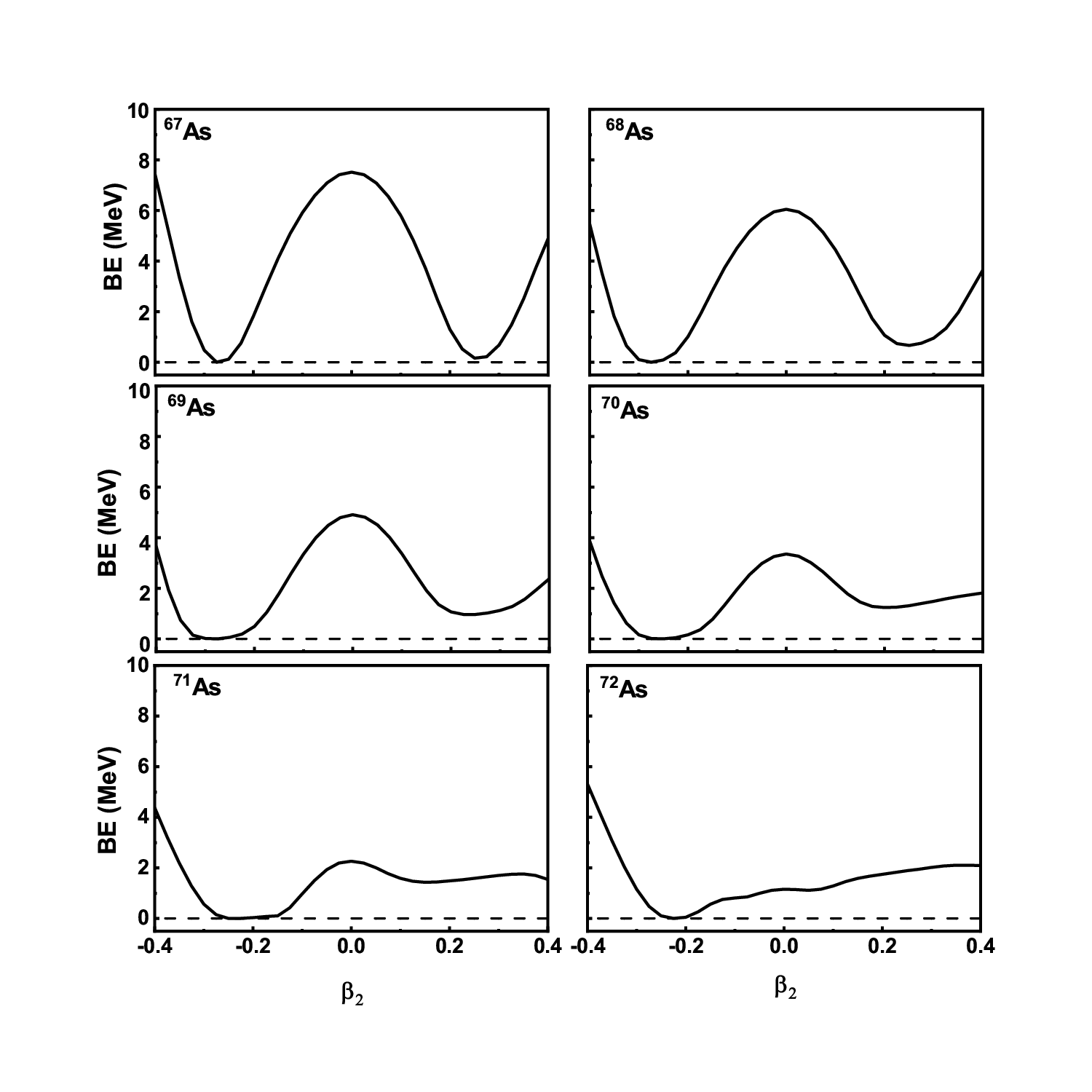}
	\caption{Computed  PECs for $^{67-72}$As by employing the DD-ME2 functional.}
	\label{Fig. 1}
\end{figure}

\begin{figure}[H]
	\centering
	\includegraphics[width=13cm]{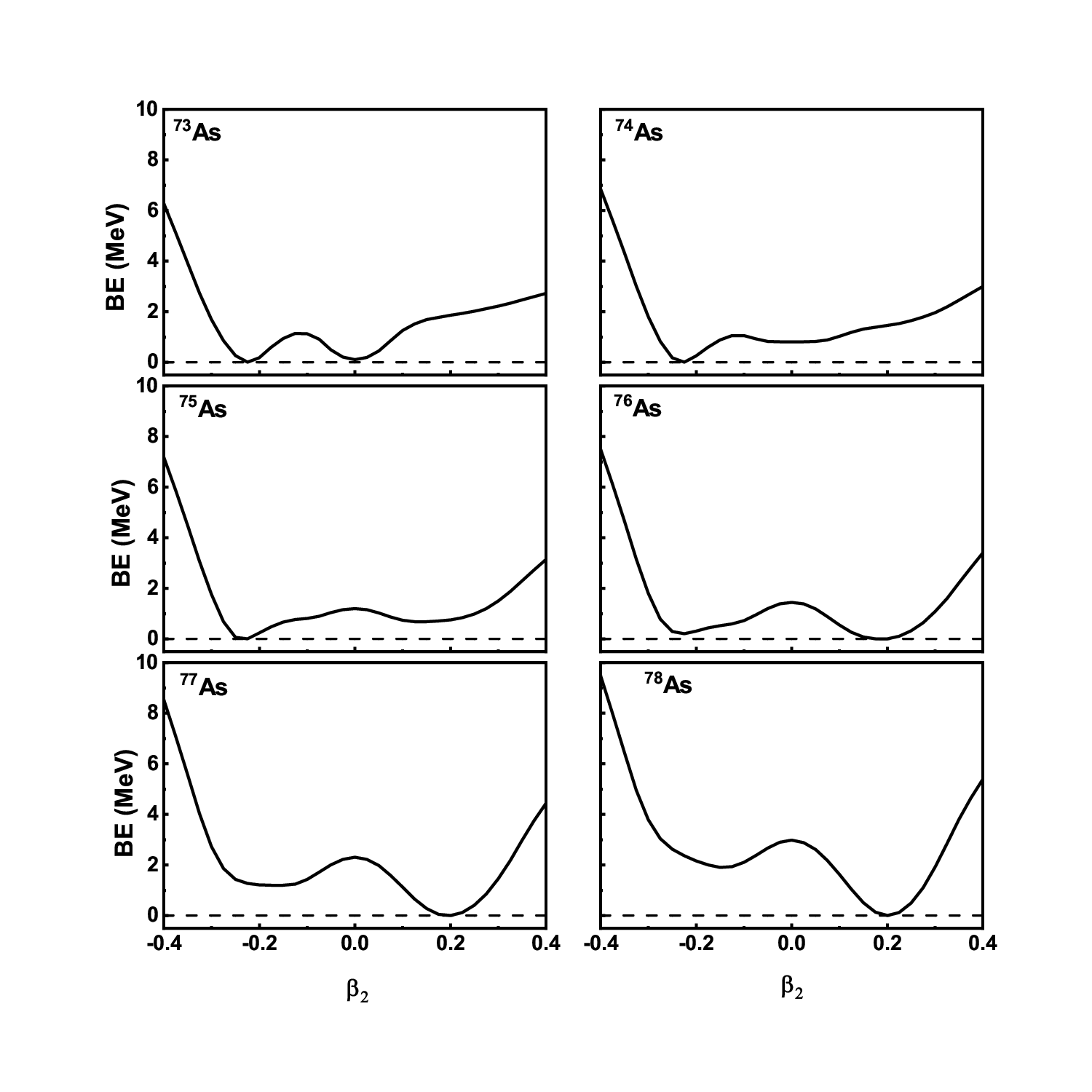}
	\caption{Same as Fig. \ref{Fig. 1} for $^{73-78}$As.}
	\label{Fig. 2}
\end{figure}

\begin{figure}[H]
	\centering
	\includegraphics[width=13cm]{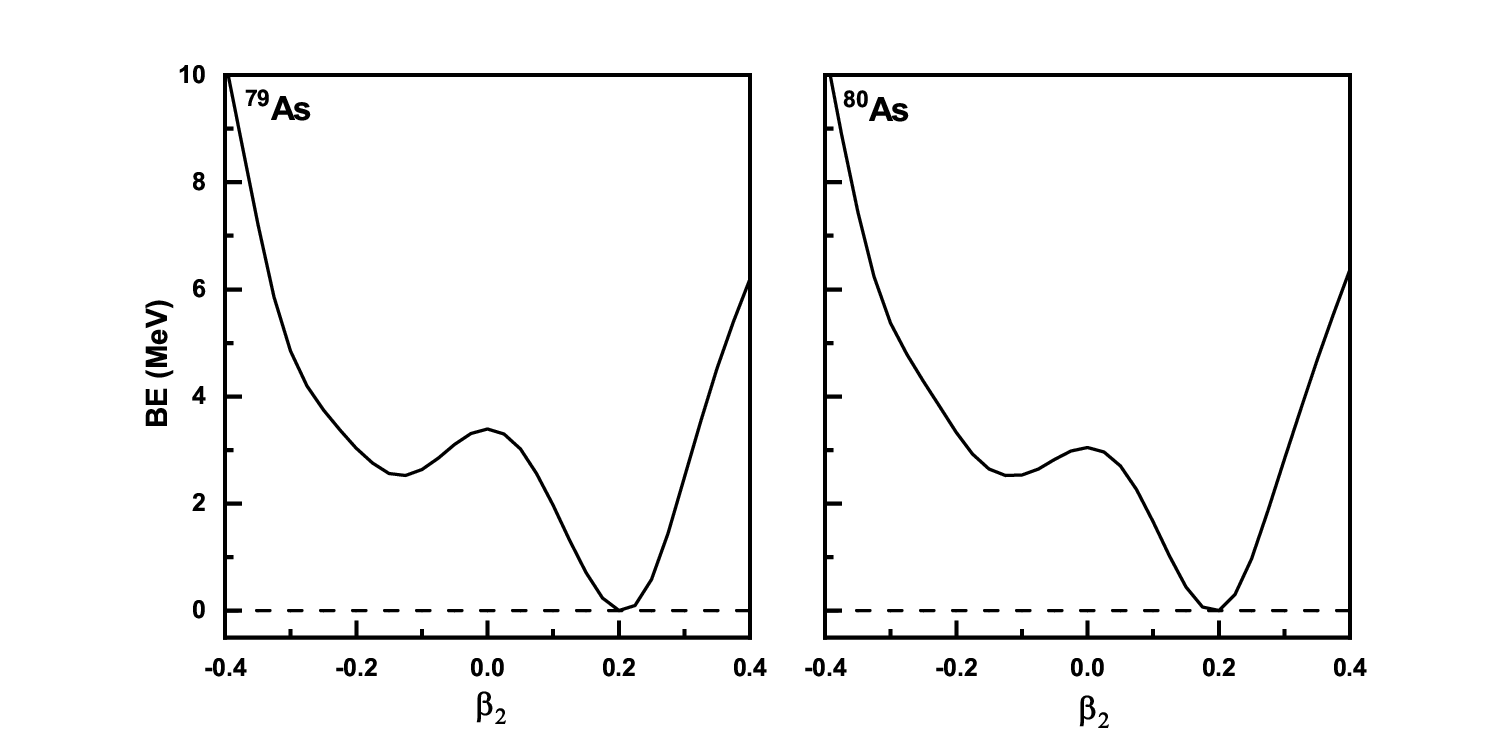}
	\caption{Same as Fig. \ref{Fig. 1} for $^{79-80}$As.}
	\label{Fig. 3}
\end{figure}

We first discuss the nuclear structure properties of As isotopes by utilizing the DD-ME2 interaction parameters in the RMF framework.  We have investigated the $\beta_{2}$ for these $^{67-80}$As nuclei in depth. To accomplish this, we performed PEC calculations of the chosen nuclei within the RMF framework. We implemented constraints on the quadrupole moment in order to compute binding energy for the analysis of PECs.
In Figs. (\ref{Fig. 1}-\ref{Fig. 3}) the PECs are expressed as a function of  $\beta_2$ for the selected As isotopes. 
For the analysis of PECs, the lowest
binding energy was used as reference. The PECs were obtained by the differences between calculated binding energy for specific $\beta_{2}$ value and the reference binding energy for the nuclei under investigation. Nuclear shapes were determined by the PEC minima. Prolate nuclei resulted from a PEC minimum located on the positive side of $\beta_{2}$, whereas oblate nuclei were found on the negative side of $\beta_{2}$. In our present case, the geometrical shapes of $Z$\,=\,33 nuclei varied. For isotopes in the mass range $A$\,=\,67-75, oblate shapes were predicted while for $A$\,=\,76-80 the DD-ME2 interaction computed prolate configurations. Table~\ref{Tab 01} shows  the terrestrial decay modes, $Q$ values~\cite{Kon21}, ground-state quadrupole deformation parameters $\beta_2$ and pairing gap values for the selected As nuclei.  The $Q$-values decrease for odd-A and odd-odd isotopes for  $\beta^{+}$ decay with increasing mass number. Beyond the stable $^{75}$As nucleus, the $Q$-values start increasing, separately for the odd-A and odd-odd isotopes of As, in the $\beta^{-}$ direction. The nuclear deformations computed from the RMF models were used as an input parameter in the pn-QRPA model to perform self-consistent calculations of the $\beta$-decay properties.
\begin{table}[H]
	\centering
	\caption{Terrestrial decay modes and values of parameters of the pn-QRPA model for selected As isotopes. } 
	\label{Tab 01}
	\scalebox{.8}{
		\begin{tabular}{c c c c c c c c c c}
			
			\hline
			Nuclei & Decay Mode & $Q_{\beta}$~(MeV) & $\beta_{2}$& $\Delta_{pp}$ (MeV) & $\Delta_{nn}$ (MeV) \\  \hline \\   
			$^{67}$As & $\beta^{+}$ & 6.0868 & -0.4226 & 1.6478 &  0.4330  \\  
			$^{68}$As & $\beta^{+}$ & 8.0843 & -0.4248 & 1.0123 & 1.0407 \\  
			$^{69}$As & $\beta^{+}$ & 3.9907 & -0.3104 & 1.6642 &  1.2249   \\  
			$^{70}$As & $\beta^{+}$ & 6.2280 & -0.3899 & 1.0901 & 1.3316 \\  
			$^{71}$As & $\beta^{+}$ & 2.0137 & -0.3526 & 1.6368 &   1.3892  \\  
			$^{72}$As & $\beta^{+}$ & 4.3559 & -0.3542 & 1.0867 &   1.4020  \\  
			$^{73}$As & EC & 0.3445 & -0.3559 & 1.7432 &  1.3004  \\  
			$^{74}$As & $\beta^{+}$ & 2.5623 & -0.3575 & 1.2233 &  1.2707 \\  
			$^{75}$As & Stable & -0.8647 & -0.3591 & 1.6793
			& 1.2959  \\  
			$^{76}$As & $\beta^{-}$ &  2.9605 & 0.2805& 1.3112 &   1.3211   \\  
			$^{77}$As & $\beta^{-}$ &  0.6832 & 0.3220 & 1.6137 &  1.2729  \\  
			$^{78}$As & $\beta^{-}$ &  4.2089 & 0.3234 & 1.2021 &  1.1606 \\  
			$^{79}$As & $\beta^{-}$ &  2.2814 & 0.3247 & 1.6112 &   1.0395 \\  
			$^{80}$As & $\beta^{-}$ &  5.5445 & 0.3261 & 1.1574 
			&   0.9948 \\  
			\hline
	\end{tabular}}
\end{table}

\begin{figure}[H]
	\centering
	\includegraphics[width=15cm]{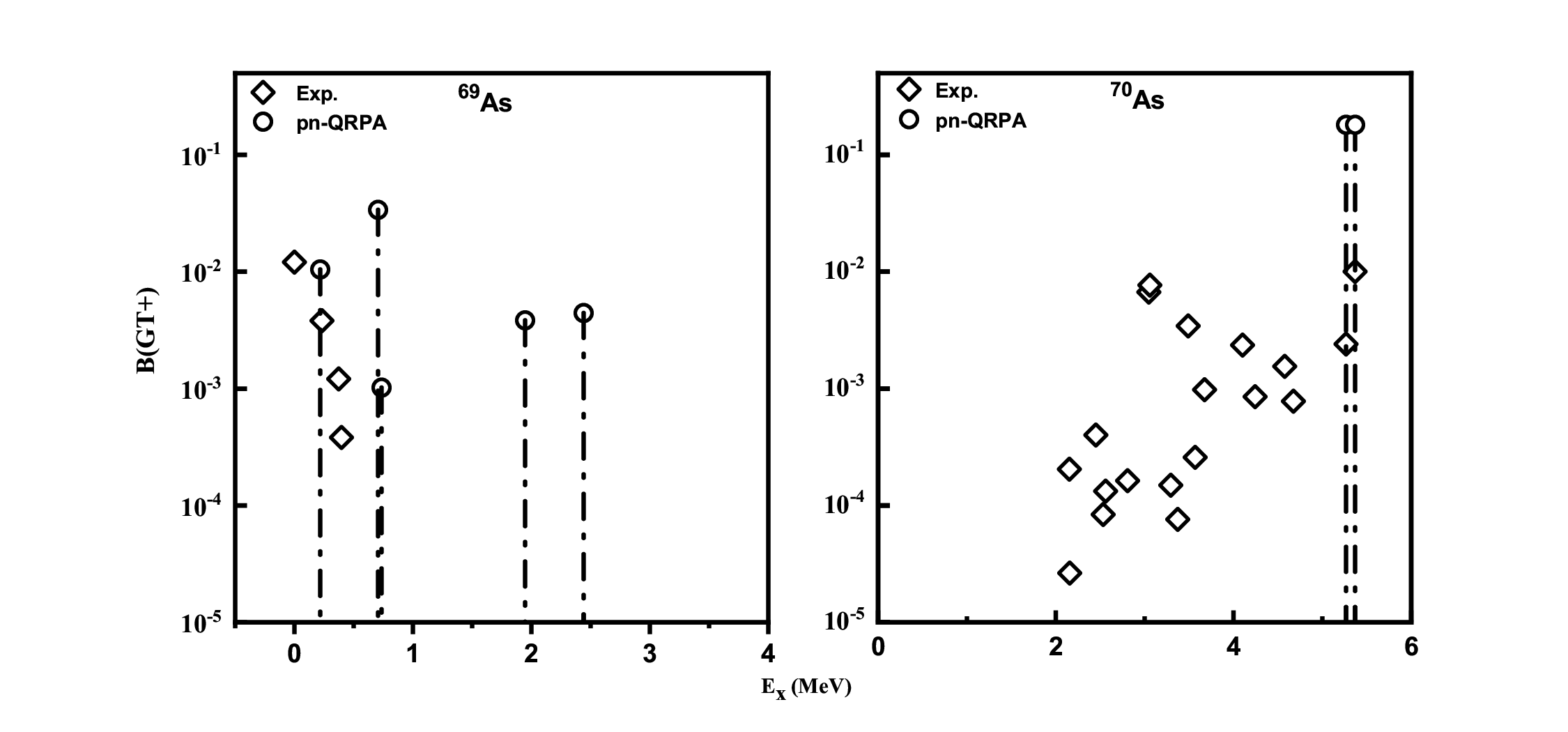}
	\caption{Comparison of the pn-QRPA calculated GT strength distributions of $^{69}$As and $^{70}$As  with experimental data~\cite{Mus70,Yan02}, respectively. The abscissa shows daughter excitation energy within the $Q$-window. }
	\label{Fig. 4}
\end{figure}

\begin{figure}[H]
	\centering
	\includegraphics[width=13cm]{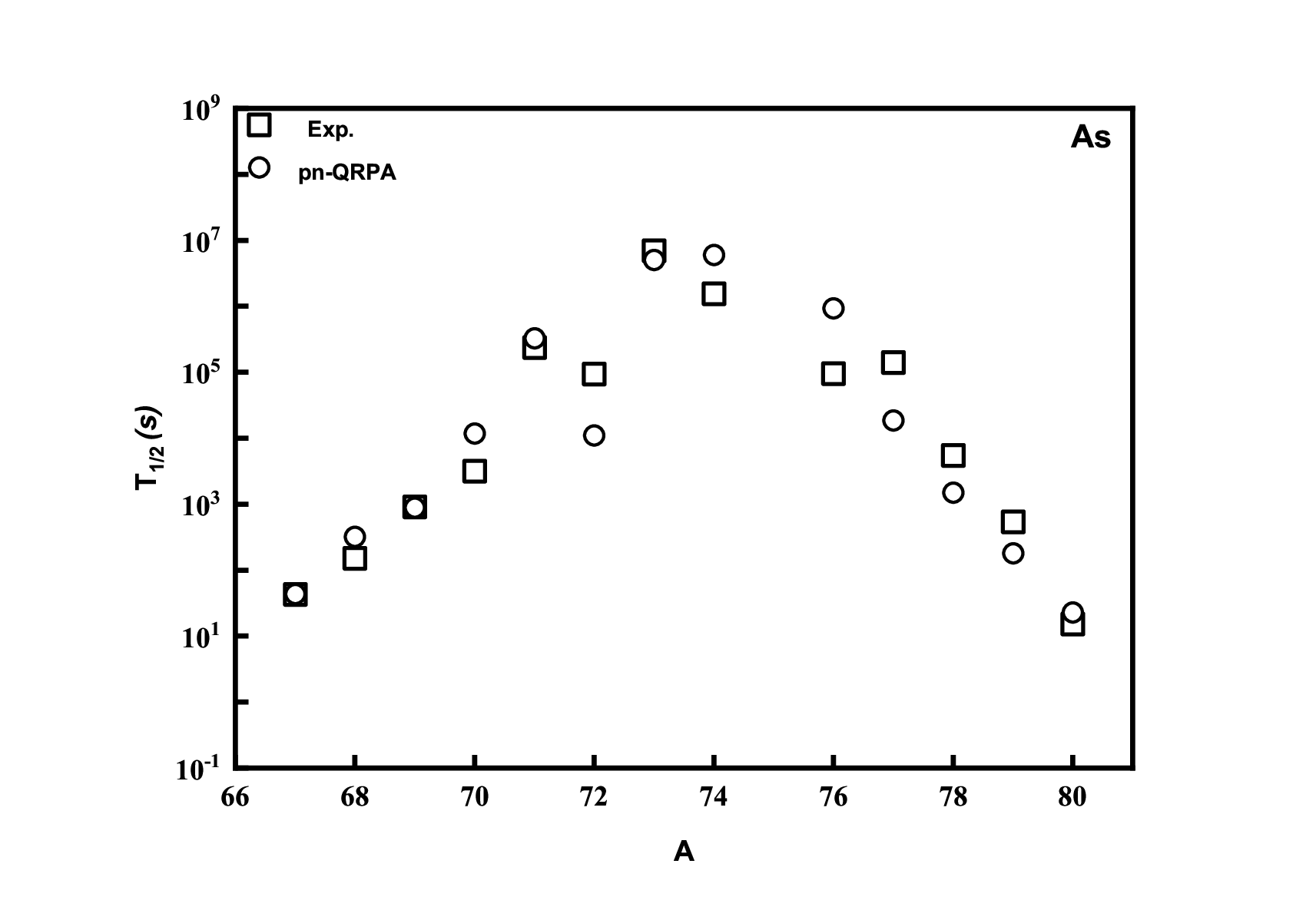}
	\caption{Comparison of the pn-QRPA calculated half-lives of As isotopes with experimental data~\cite{Kon21}.}
	\label{Fig. 5}
\end{figure}

\begin{figure}[H]
	\centering
	\includegraphics[width=13cm]{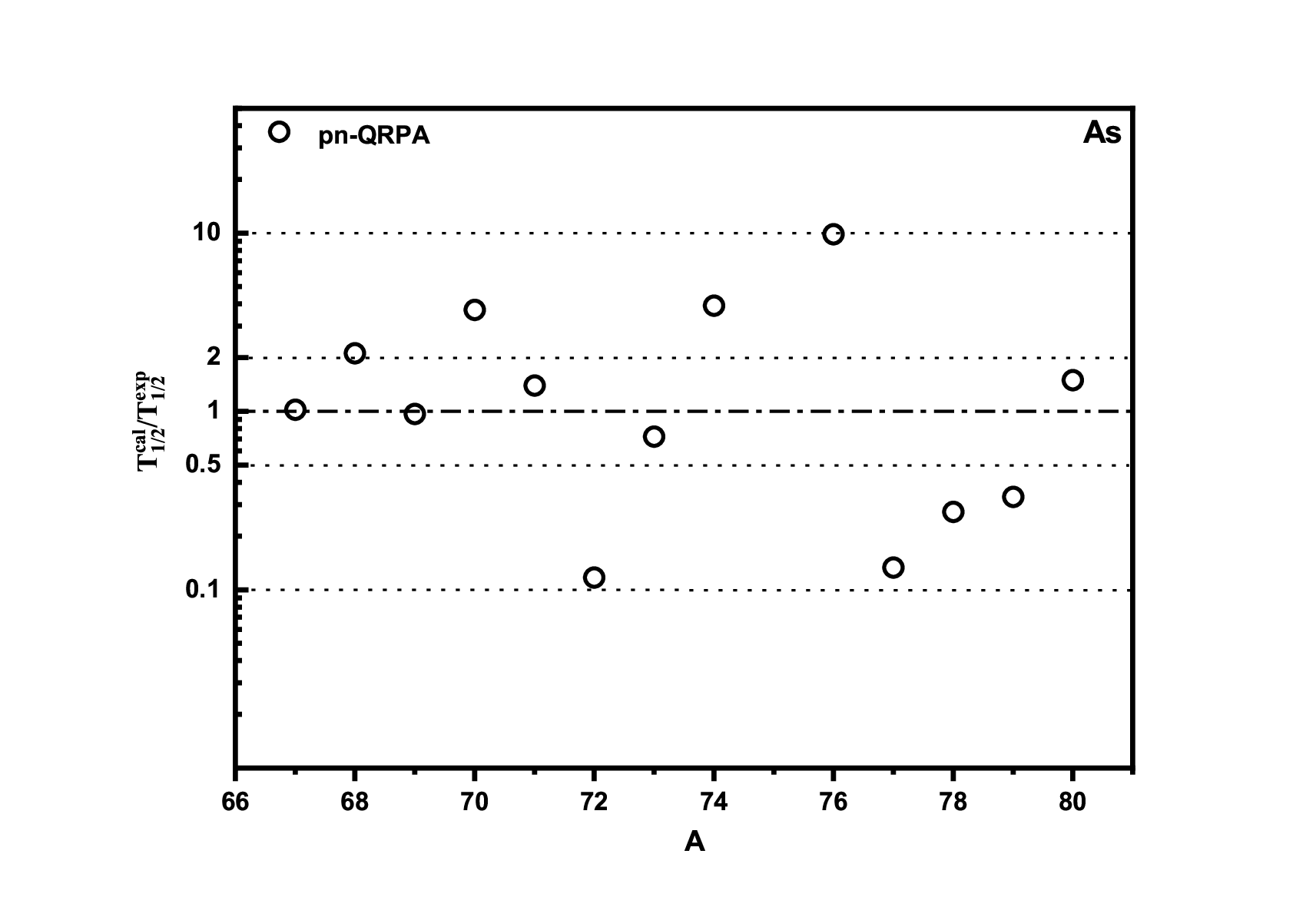}
	\caption{Ratio of the calculated to measured~\cite{Kon21}  half-lives for As isotopes.}
	\label{Fig. 6}
\end{figure}

Fig.~(\ref{Fig. 4}) displays the comparison between the measured and calculated  GT$+$ strength distributions for $^{69, 70}$As. The experimental data, taken from Refs.~\cite{Mus70,Yan02}, was reported up to  excitation energies of 0.39~MeV and 5.37~MeV for $^{69}$As and $^{70}$As, respectively.  The calculated GT distributions are shown within the $Q$-window  (3.99~MeV for $^{69}$As and 6.23~MeV for $^{70}$As). The low-lying transitions are in good comparison with the measured data for $^{69}$As decay. For the decay of $^{70}$As, the measured strength is distributed  between (2~-~5.5)~MeV. The  pn-QRPA missed the low-lying transitions. The calculated GT distributions are not much fragmented for the case of $^{70}$As. The current pn-QRPA model considers $1p-1h$ correlations. Inclusion of $2p-2h$ correlations may improve the calculation and would be taken as a future assignment. 
 The comparison of calculated and measured~\cite{Kon21} half-lives is shown in  Fig.~(\ref{Fig. 5}) for the selected As nuclei.  Fig.~(\ref{Fig. 6}) displays the ratios of pn-QRPA calculated to the measured half-lives. For all cases under investigation, the calculated half-lives were reproduced within a factor of 10 of the measured ones. This indicates good accuracy of the underlying nuclear model which we attribute to the optimal choice of model parameters. The pn-QRPA model is known to have a good prediction power for $\beta$-decay half-lives of neutron-rich nuclei~\cite{Hir93,Hom96}.
	
Table \ref{Tab 02} shows the comparison of our calculated log $\textit{ft}$ values using the pn-QRPA model for $\beta^{+}$/$EC$ decays of odd-A As nuclei with measured data and previous calculations. The proton-neutron interacting boson-fermion model (IBFM-2) was used to calculate the state-by-state transitions of selected odd-A As nuclei~\cite{ Bra04}. Later, the same nuclear model was used to perform constrained self-consistent
mean-field (SCMF) calculations based on a universal
energy density functional and a pairing interaction~\cite{Nom22}. This calculation is referred to as IBFM-2-SCMF in Table \ref{Tab 02}. The measured log $\textit{ft}$ values  were taken from Ref.~\cite{NNDC}.  For majority of the transitions, the IBFM-2 calculated values
are smaller than the experimental data. The nuclear matrix elements were overestimated because of admixture of other components including intruder states in the wavefunctions of the nuclear model~\cite{Bra04}.  The log $\textit{ft}$ values, sensitive to the wave functions for the
initial and final states, computed by the IBFM-2-SCMF model were systematically larger than the IBFM-2 values.  The IBFM-2 framework was largely based on  empirical parameters for the even-even
IBM-2 core Hamiltonian. Consequently, phenomenological single-particle energies were adopted in IBFM-2 calculation. On the other hand, most
of the parameters were determined based on the energy density functional
calculations in the IBFM-2-SCMF model. It is noted that our self-consistent pn-QRPA calculations are in good agreement with the measured data.
\begin{table}[H]
	\centering
	\caption{ Comparison of the calculated and measured~\cite{NNDC} log $\textit{ft}$ values for $\beta^{+}$/$EC$ decays of odd-A As nuclei. The last three columns show the theoretical values calculated using the pn-QRPA, IBFM-2~\cite{Bra04} and IBFM-2-SCMF~\cite{Nom22} models, respectively.}
	\label{Tab 02}
	\addtolength{\tabcolsep}{1pt}
	\scalebox{0.6}{
		\begin{tabular}{ccccccc}
			\hline
			\multicolumn{6}{c}{log $\textit{ft}$} \\
			\hline
			Decay & I$_{i} \rightarrow $I$_{f}$ & Exp. & pn-QRPA & IBFM-2  & IBFM-2-SCMF \\
			\hline
			$^{67}$As $\rightarrow$ $^{67}$Ge & $5/2^{-}_1 \rightarrow 5/2^{-}_1$ & 5.44 & 5.08 & - & 4.15\\
			& $5/2^{-}_1 \rightarrow 5/2^{-}_2$ & 5.92 & 6.17 & -  & 6.63 \\
			& $5/2^{-}_1 \rightarrow 5/2^{-}_3$ & 6.40 & 5.25 & -  & 6.08 \\
			& $5/2^{-}_1 \rightarrow 3/2^{-}_1$ & 6.18  & 5.27 & - & 6.49 \\
			& $5/2^{-}_1 \rightarrow 3/2^{-}_2$ & 5.64 & 7.35 & -  & 7.61 \\
			$^{69}$As $\rightarrow$ $^{69}$Ge & $5/2^{-}_1 \rightarrow 5/2^{-}_1$ & 5.49 & 5.44 & 4.26 & 4.77 \\
			& $5/2^{-}_1 \rightarrow 5/2^{-}_2$ & 6.94  & 6.89 & 6.65 & 6.92 \\
			& $5/2^{-}_1 \rightarrow 5/2^{-}_3$ & 6.80  & 6.76 & 5.33  & 5.63 \\
			& $5/2^{-}_1 \rightarrow 5/2^{-}_4$ & 6.47 & 6.42 & 5.49 & 5.98  \\
			& $5/2^{-}_1 \rightarrow 5/2^{-}_5$ & 5.95 & 5.90 & - & 7.15 \\
			& $5/2^{-}_1 \rightarrow 3/2^{-}_1$ & 6.05 & 6.00 & 5.88 & 7.58 \\
			& $5/2^{-}_1 \rightarrow 3/2^{-}_2$ & 7.21 & 7.15 & 7.90 & 7.44 \\
			& $5/2^{-}_1 \rightarrow 3/2^{-}_3$ & 6.71  & 6.66 & 5.07 & 6.43 \\
			& $5/2^{-}_1 \rightarrow 3/2^{-}_4$ & 5.82 & 5.77 & 6.46 & 7.07 \\
			& $5/2^{-}_1 \rightarrow 3/2^{-}_5$ & 6.21 & 6.16 & 6.73 & 8.00 \\
			& $5/2^{-}_1 \rightarrow 7/2^{-}_1$ & 6.20 & 6.15 & 7.54 & 10.85 \\
			& $5/2^{-}_1 \rightarrow 7/2^{-}_2$ & - & 5.39 & - & 7.46 \\
			$^{71}$As $\rightarrow$ $^{71}$Ge & $5/2^{-}_1 \rightarrow 5/2^{-}_1$ &5.85 & 5.97 &4.60 & 5.92 \\
			& $5/2^{-}_1 \rightarrow 5/2^{-}_2$ & - & 6.40 & 6.08 & 6.28 \\
			& $5/2^{-}_1 \rightarrow 5/2^{-}_3$ & 6.86  & 6.99 & 5.63 & 6.55 \\
			& $5/2^{-}_1 \rightarrow 5/2^{-}_4$ & 9.14  & 7.58 & 5.55 & 7.74 \\
			& $5/2^{-}_1 \rightarrow 5/2^{-}_5$ & 6.96  & 7.30 & - & 6.84 \\
			& $5/2^{-}_1 \rightarrow 3/2^{-}_1$ & 7.19  & 7.30 & 6.52 & 6.74 \\
			& $5/2^{-}_1 \rightarrow 3/2^{-}_2$ & >8.60  & 8.70 & 7.79 & 7.47 \\
			& $5/2^{-}_1 \rightarrow 3/2^{-}_3$ & 6.33  & 6.45 & 5.73 & 7.24 \\
			& $5/2^{-}_1 \rightarrow 3/2^{-}_4$ & 7.43  & 7.55 & 5.21 & 8.25 \\
			& $5/2^{-}_1 \rightarrow 3/2^{-}_5$ & 6.94  & 7.07 & 7.34 & 8.10 \\
			& $5/2^{-}_1 \rightarrow 7/2^{-}_1$ & 8.79  & 8.70 & 7.6 & 8.38 \\
			& $5/2^{-}_1 \rightarrow 7/2^{-}_2$ & 7.29  & 7.42 & -  & 7.85 \\
			$^{73}$As $\rightarrow$ $^{73}$Ge & $3/2^{-}_1 \rightarrow 1/2^{-}_1$ & 5.40 & 5.47 & - & 3.82  \\
			\hline
		\end{tabular}
	}
\end{table}
Comparison of the pn-QRPA calculated log $\textit{ft}$ values with measured data~\cite{NNDC} and IBFM-2-SCMF model for the case of odd-odd As isotopes is shown in Table~\ref{Tab 03}. The IBFM-2 model was not capable of computing $\textit{ft}$ values for odd-odd nuclei. The pn-QRPA calculated log $\textit{ft}$ values are in better agreement with the measured data for $^{68}$As. For the decay $^{70}$As $\rightarrow$ $^{70}$Ge, our calculated log $\textit{ft}$ values are smaller than the measured ones. The short-coming of our model, for the case of $^{70}$As, resulting in lesser fragmentation was discussed earlier.

\begin{table}[H]
	\centering
	\caption{Comparison of the calculated and measured ~\cite{NNDC} log $\textit{ft}$ values for $\beta^{+}$/$EC$ on odd-odd As nuclei. The last two columns show the theoretical values calculated using the pn-QRPA and IBFM-2-SCMF~\cite{Nom22} models, respectively.}
	\label{Tab 03}
	\addtolength{\tabcolsep}{1pt}
	\scalebox{.6}{\begin{tabular}{cccccc}
			\hline
			\multicolumn{5}{c}{log $\textit{ft}$} \\
			\hline
			Decay & I$_{i} \rightarrow $I$_{f}$ & Exp. & pn-QRPA & IBFM-2-SCMF \\
			\hline
			
			$^{68}$As→$^{68}$Ge &  $3^+_1 \rightarrow 2^+_1$ &  7.38 & 7.02 & 6.66    \\
			& $3^+_1 \rightarrow 2^+_2$ & 6.86  & 6.72 & 6.95    \\
			& $3^+_1 \rightarrow 2^+_3$ & 6.89 & 6.74 & 6.34    \\
			& $3^+_1 \rightarrow 2^+_4$ &  7.24 & 5.20  & 5.81    \\
			& $3^+_1 \rightarrow 2^+_5$ &  6.57  &  5.19  & 7.21    \\
			&  $3^+_1 \rightarrow 4^+_1$ & 7.02  & 6.89 & 6.34    \\
			&  $3^+_1 \rightarrow 4^+_2$ & 6.74  & 5.97 & 5.73    \\
			&  $3^+_1 \rightarrow 4^+_3$ & 5.97  & 5.49 & 6.63  \\
			$^{70}$As→$^{70}$Ge & $4^+_1 \rightarrow 4^+_1$ &  7.30      &  4.77 & 6.58     \\
			&$ 4^+_1 \rightarrow 4^+_2$ &  7.37 & 4.77 & 6.03   \\
			& $4^+_1 \rightarrow 4^+_3$ &   5.69 & 4.50 & 6.01    \\
			& $4^+_1 \rightarrow 3^+_1$ &   6.97 & 4.53  & 10.74  \\
			\hline
	\end{tabular}}
\end{table}

The time derivative of the lepton-to-baryon fraction $Y_e$ is a key parameter to be monitored during the presupernova evolution of a massive star. For a particular nucleus it is given by
\begin{equation}
	\dot{Y}_{e}^{(\beta^{-}+PC)} = \frac{\tau}{A}\lambda^{(\beta^{-}+PC)} , \nonumber
\end{equation}
\begin{equation}
	\dot{Y}_{e}^{(\beta^{+}+EC)} = -\frac{\tau}{A}\lambda^{(\beta^{+}+EC)} , \nonumber
\end{equation}
where $\tau$ is the abundance of the nucleus and $A$ is the mass number. It is noted that sum of $\beta^{+}$ and $EC$ have a negative contribution whereas sum of $\beta^{-}$ and $PC$ rates have a positive contribution to $\dot{Y}_{e}$.  We have calculated the $\beta^{\pm}$, $EC$ and $PC$ rates of selected As isotopes in stellar matter using the pn-QRPA model with nuclear deformations computed from the RMF model. Our calculation of stellar rates is fully microscopic in nature. We did not use the so-called Brink-Axel hypothesis~\cite{Ike64} for calculation of excited states GT strength distributions. Earlier these stellar  rates were computed using the independent-particle model (IPM) with the then available experimental data  for free nucleons and 226 nuclei in the mass region $A$ = 21--60~\cite{Ful82}. Later, Pruet and Fuller using the same model,   updated these calculations for  the mass region $A$ = 65--80~\cite{Pre03}.  In the revised calculation, the authors undertook corrections in the  placement of GT centroids by Ref.~\cite{Ful82} and introduced a better treatment of high-temperature partition functions. The calculations assumed Brink-Axel hypothesis and will be referred to as IPM-03 here onward.  We present the calculation of pn-QRPA and IPM-03 stellar rates  in  Tables (\ref{Tab 04}~--~\ref{Tab 05}). Here we show the sum of $\beta^{+}$ and $EC$ in one direction and sum of $\beta^{-}$ and $PC$ rates in the other direction, for the selected As isotopes. Tables (\ref{Tab 04}~--~\ref{Tab 05}) also compare the pn-QRPA stellar rates with the IPM-03 calculations at selected temperature (3, 10, 30) GK and density ($10^3, 10^7$, $10^{11}$)~g/cm$^3$ values. The IPM-03 method applied a quenching factor of 3 and 4 for ($\beta^{-}$+ $PC$) and  ($\beta^{+}$+ $EC$) transitions, respectively. No explicit quenching factor was included in pn-QRPA calculations. The stellar rates increase with soaring core temperatures. This occurs because the occupation probability of parent states rises with higher temperatures, leading to a finite contribution to the total stellar rates. As the density of the stellar core increases by orders of magnitude,  the ($\beta^{+}$+ $EC$)  rates  increase  due to a corresponding elevation in electron chemical potential. Conversely, the ($\beta^{-}$+ $PC$) rates decrease with increasing core densities due to a substantial reduction in the available phase space (the degeneracy parameter is negative for positrons). The ($\beta^{+}$+ $EC$)  rates become meaningful only at high temperatures and densities where the pn-QRPA rates are much bigger (up to a factor 33 or more) than the IPM-03 rates. Table \ref{Tab 05} on the other hand shows that the ($\beta^{-}$+ $PC$) rates have a finite contribution (important for stellar evolution) only at high temperature and low density values. This is due to the Pauli blocking effect at high densities as well as considerable reduction in phase space discussed earlier. Once again we note that the IPM-03 rates are smaller than the pn-QRPA rates. The sum of partition function, in the IPM-03 computation, had a large number of states without associated weak interaction strength. Treatment of partition functions and quenching of GT strength in IPM-03 calculation could be the probable reason for their smaller rates.                

\begin{table}[H]
	\centering
	\caption{Calculated sum of $\beta^{+}$ and $EC$ stellar rates for As isotopes as a function of core temperature and density values in units of s$^{-1}$.  The temperatures ($T_9$) and densities ($\rho$) are given in units of  $10^9$ K and g/cm$^3$, respectively.  Calculated rates are compared with the previous IPM-03 calculation~\cite{Pre03}.}
	\label{Tab 04}
	\addtolength{\tabcolsep}{1pt}
	\scalebox{.7}{
		\begin{tabular}{ccccccccc}
			\toprule
			\multicolumn{1}{c}{}\\
			\multicolumn{2}{c}{} & \multicolumn{2}{c}{{$^{67}$As}} & \multicolumn{2}{c}{{$^{68}$As}} & \multicolumn{2}{c}{{$^{69}$As}}\\
			\cmidrule(rl){3-4} \cmidrule(rl){5-6} \cmidrule(rl){7-8}
			$\rho$ & T$_9$  & $\lambda_{pn-QRPA}^{(\beta^{+}+EC)}$ & $\lambda_{IPM-03}^{(\beta^{+}+EC)}$ & $\lambda_{pn-QRPA}^{(\beta^{+}+EC)}$ & $\lambda_{IPM-03}^{(\beta^{+}+EC)}$  & $\lambda_{pn-QRPA}^{(\beta^{+}+EC)}$ & $\lambda_{IPM-03}^{(\beta^{+}+EC)}$ \\
			\midrule
			$10^{3}$  & $3$ &	$3.75\times 10^{-02}$ &	$6.49\times 10^{-02}$  & $3.23 \times 10^{-01}$ & $5.37 \times 10^{-02}$ & $4.60\times 10^{-03}$  &	$8.61\times 10^{-03}$
			\\
			& $10$ &	$1.58\times 10^{+00}$ &	$8.50\times 10^{-01}$
			 & $4.81 \times 10^{+00}$ & $7.10 \times 10^{-01}$ &	$3.54\times 10^{-01}$ &	$4.19\times 10^{-01}$
			\\
			& $30$ &	$8.58\times 10^{+02}$ &	$2.48\times 10^{+02}$ & $1.76 \times 10^{+03}$ & $4.46 \times 10^{+02}$  &	$4.05\times 10^{+02}$ &	$1.84\times 10^{+02}$
			\\
			$10^{7}$ & $3$ &	$9.59\times 10^{-02}$ &	$9.39\times 10^{+02}$  & $7.64 \times 10^{-01}$ & $1.02\times 10^{-01}$ &	$2.81\times 10^{-02}$ &	$4.44\times 10^{-02}$
			\\
			& $10$ &	$1.93\times 10^{+00}$ &	$1.01\times 10^{+00}$  & $5.82 \times 10^{+00}$ & $8.31 \times 10^{-01}$  &	$4.37\times 10^{-01}$ &	$5.15\times 10^{-01}$
			\\
			& $30$ &	$8.64\times 10^{+02}$ & $2.50\times 10^{+02}$  & $1.77 \times 10^{+03}$ & $4.49 \times 10^{+02}$ &	$4.07\times 10^{+02}$ &	$1.85\times 10^{+02}$
			\\
			$10^{11}$  & $3$ &	$9.51\times 10^{+04}$ &	$6.43\times 10^{+04}$ & $2.61 \times 10^{+05}$ & $5.45 \times 10^{+04}$ &	$8.26\times 10^{+04}$ &	$4.80\times 10^{+04}$
			\\
			& $10$ &	$2.52\times 10^{+05}$ &	$6.85\times 10^{+04}$ & $4.57 \times 10^{+05}$ & $5.71 \times 10^{+04}$ &	$1.11\times 10^{+05}$ &	 $4.98\times 10^{+04}$
			\\
			& $30$  &	$6.44\times 10^{+05}$	& $1.13\times 10^{+05}$ & $1.15 \times 10^{+06}$ & $1.77 \times 10^{+05}$ &	$3.43\times 10^{+05}$	& $8.67\times 10^{+04}$
			\\
			\bottomrule
		\end{tabular}	}
	\addtolength{\tabcolsep}{1pt}
	\scalebox{.7}{
		\begin{tabular}{ccccccccc}
			\multicolumn{1}{c}{} \\
			\multicolumn{2}{c}{} & \multicolumn{2}{c}{{$^{70}$As}} & \multicolumn{2}{c}{{$^{71}$As}} & \multicolumn{2}{c}{{$^{72}$As}}\\
			\cmidrule(rl){3-4} \cmidrule(rl){5-6} \cmidrule(rl){7-8}
			$\rho$ & T$_9$  & $\lambda_{pn-QRPA}^{(\beta^{+}+ EC)}$ & $\lambda_{IPM-03}^{(\beta^{+}+EC)}$  & $\lambda_{pn-QRPA}^{(\beta^{+}+ EC)}$ & $\lambda_{IPM-03}^{(\beta^{+}+EC)}$ & $\lambda_{pn-QRPA}^{(\beta^{+}+ EC)}$ & $\lambda_{IPM-03}^{(\beta^{+}+ EC)}$ \\
			\midrule
			$10^{3}$  & $3$
			& $2.54 \times 10^{-02}$ & $8.48 \times 10^{-03}$ &	$2.58\times 10^{-04}$ &	$3.63\times 10^{-04}$ & $5.57 \times 10^{-04}$ & $8.60 \times 10^{-04}$ \\
			& $10$ & $1.11 \times 10^{+00}$ & $3.12 \times 10^{-01}$ & $1.10\times 10^{-01}$ &	$1.81\times 10^{-01}$ & $3.46 \times 10^{-01}$ & $1.35 \times 10^{-01}$ \\
			& $30$ & $7.54 \times 10^{+02}$ & $3.17 \times 10^{+02}$ &	$2.58\times 10^{+02}$ &	$1.01\times 10^{+02}$ & $5.19 \times 10^{+02}$ & $2.02 \times 10^{+02}$ \\
			$10^{7}$  & $3$ & $1.44 \times 10^{-01}$ & $3.98 \times 10^{-02}$ &	$6.48\times 10^{-03}$&	$1.18\times 10^{-02}$ & $1.21 \times 10^{-02}$ & $1.21 \times 10^{-02}$ \\
			& $10$ & $1.38 \times 10^{+00}$ & $4.29 \times 10^{-04}$ &	$1.37\times 10^{-01}$ &	$2.23\times 10^{-01}$ & $4.31 \times 10^{-01}$ & $1.67 \times 10^{-01}$  \\
			& $30$  & $7.59 \times 10^{+02}$ & $3.19 \times 10^{+02}$ &	$2.59\times 10^{+02}$ &	$1.02\times 10^{+02}$ & $5.24 \times 10^{+02}$ & $2.04 \times 10^{+02}$ \\
			$10^{11}$  & $3$ & $2.00 \times 10^{+05}$ & $3.54 \times 10^{+04}$  &	$5.40\times 10^{+04}$ &	$2.34\times 10^{+04}$ & $1.33 \times 10^{+05}$ & $1.78 \times 10^{+04}$ \\
			& $10$ & $2.86 \times 10^{+05}$ & $3.61 \times 10^{+04}$ &	$7.52\times 10^{+04}$ &	$2.46\times 10^{+04}$ & $2.19 \times 10^{+05}$ & $1.82 \times 10^{+04}$ \\
			& $30$ & $6.46 \times 10^{+05}$ & $1.27 \times 10^{+05}$ &	$2.82\times 10^{+05}$ & $4.91\times 10^{+04}$
 & $5.64 \times 10^{+05}$ & $8.07 \times 10^{+04}$ \\
			\bottomrule
		\end{tabular}	}
	\addtolength{\tabcolsep}{1pt}
	\scalebox{.7}{
		\begin{tabular}{ccccccccc}
			\multicolumn{1}{c}{} \\
			\multicolumn{2}{c}{} & \multicolumn{2}{c}{{$^{73}$As}} & \multicolumn{2}{c}{{$^{74}$As}} & \multicolumn{2}{c}{{$^{75}$As}}\\
			\cmidrule(rl){3-4} \cmidrule(rl){5-6} \cmidrule(rl){7-8}
			$\rho$ & T$_9$ & $\lambda_{pn-QRPA}^{(\beta^{+}+ EC)}$ & $\lambda_{IPM-03}^{(\beta^{+}+EC)}$ & $\lambda_{pn-QRPA}^{(\beta^{+}+ EC)}$ & $\lambda_{IPM-03}^{(\beta^{+}+EC)}$ & $\lambda_{pn-QRPA}^{(\beta^{+}+ EC)}$ & $\lambda_{IPM-03}^{(\beta^{+}+ EC)}$ \\
			\midrule
			$10^{3}$  & $3$ &	$2.81\times 10^{-05}$ &	$3.52\times 10^{+04}$ & $7.39 \times 10^{-05}$ & $2.86 \times 10^{-04}$ &	$3.57\times 10^{-07}$ &	$3.29\times 10^{-07}$
			\\
			& $10$  &	$5.03\times 10^{-02}$ &	$4.84\times 10^{-02}$ & $1.81 \times 10^{-01}$ & $1.34 \times 10^{-01}$ &	$ 1.21\times 10^{-02}$ &	$1.46\times 10^{-02}$
			\\
			& $30$ &	$2.70\times 10^{+02}$ &	$6.36\times 10^{+01}$ & $5.40 \times 10^{+02}$ & $1.46 \times 10^{+02}$ &	$1.67\times 10^{+02}$ & $4.60\times 10^{+01}$
			\\
			$10^{7}$  & $3$ &	$1.01\times 10^{-03}$ &	$1.12\times 10^{-03}$ & $2.90 \times 10^{-03}$ & $9.87 \times 10^{-03}$ & $1.82\times 10^{-05}$ & $1.67\times 10^{-05}$
			\\
			& $10$ &	$6.28\times 10^{-02}$ &	$6.02\times 10^{-02}$ & $2.26 \times 10^{-01}$ & $1.67 \times 10^{-01}$ &	 $1.52\times 10^{-02}$ &	$1.83\times 10^{-02}$
			\\
			& $30$ & $2.72\times 10^{+02}$ &	$6.40\times 10^{+01}$ & $5.45 \times 10^{+02}$ & $1.47 \times 10^{+02}$ &	$1.69\times 10^{+02}$ &	$4.63\times 10^{+01}$
			\\
			$10^{11}$  & $3$ &	$5.42\times 10^{+04}$	& $1.47\times 10^{+04}$ & $1.64 \times 10^{+05}$ & $1.87 \times 10^{+04}$  &	$4.90\times 10^{+04}$ &	$1.22\times 10^{+04}$
			\\
			& $10$  &	$9.48\times 10^{+04}$ &	$1.49\times 10^{+04}$ & $2.52 \times 10^{+05}$ & $1.91 \times 10^{+04}$ &	$8.32\times 10^{+04}$ &	$1.24\times 10^{+04}$
			\\
			& $30$  &	$3.67\times 10^{+05}$ &	$3.22\times 10^{+04}$ & $7.03 \times 10^{+05}$ & $6.21 \times 10^{+04}$ &	$2.81\times 10^{+05}$ & $2.58\times 10^{+04}$
			\\
			\bottomrule
		\end{tabular}	}
	\small
	\addtolength{\tabcolsep}{1pt}
	\scalebox{.7}{
		\begin{tabular}{ccccccccc}
			\multicolumn{1}{c}{} \\
			\multicolumn{2}{c}{} & \multicolumn{2}{c}{{$^{76}$As}} & \multicolumn{2}{c}{{$^{77}$As}} & \multicolumn{2}{c}{{$^{78}$As}}\\
			\cmidrule(rl){3-4} \cmidrule(rl){5-6} \cmidrule(rl){7-8}
			$\rho$ & T$_9$  & $\lambda_{pn-QRPA}^{(\beta^{+}+ EC)}$ & $\lambda_{IPM-03}^{(\beta^{+}+EC)}$ & $\lambda_{pn-QRPA}^{(\beta^{+}+ EC)}$ & $\lambda_{IPM-03}^{(\beta^{+}+EC)}$ & $\lambda_{pn-QRPA}^{(\beta^{+}+ EC)}$ & $\lambda_{IPM-03}^{(\beta^{+}+ EC)}$ \\
			\midrule
			$10^{3}$  & $3$ & $1.74 \times 10^{-07}$ & $1.63 \times 10^{-05}$ & $3.74\times 10^{-09}$& $1.76\times 10^{-09}$ & $2.74 \times 10^{-14}$ & $4.84 \times 10^{-07}$  \\
			& $10$  & $3.06 \times 10^{-02}$ & $5.41 \times 10^{-02}$ &	$4.40\times 10^{-03}$ &	$3.67\times 10^{-03}$  & $1.67 \times 10^{-03}$ & $1.33 \times 10^{-02}$  \\
			& $30$ & $3.41 \times 10^{+02}$ & $1.01 \times 10^{+02}$ &	$1.60\times 10^{+02}$ & $	3.16\times 10^{+01}$ & $2.87 \times 10^{+02}$ & $6.16 \times 10^{+01}$  \\
			$10^{7}$  & $3$  & $8.59 \times 10^{-06}$ & $7.43 \times 10^{-04}$  & $1.92\times 10^{-07}$ &	 $9.12\times 10^{-08}$ & $1.43 \times 10^{-12}$ & $2.43 \times 10^{-05}$  \\
			& $10$  & $3.84 \times 10^{-02}$ & $6.77 \times 10^{-02}$  &	$5.52\times 10^{-03}$ &	$4.59\times 10^{-03}$ & $2.09 \times 10^{-03}$ & $1.67 \times 10^{-02}$  \\
			& $30$  & $3.44 \times 10^{+02}$ & $1.02 \times 10^{+02}$ &	$1.61\times 10^{+02}$ & $3.18\times 10^{+01}$  & $2.89 \times 10^{+02}$ & $6.22 \times 10^{+01}$ \\
			$10^{11}$  & $3$ & $8.02 \times 10^{+04}$ & $1.59 \times 10^{+04}$  &	$4.54\times 10^{+04}$ & $9.89\times 10^{+03}$ & $8.85 \times 10^{+04}$ & $1.22 \times 10^{+04}$  \\
			& $10$  & $1.86 \times 10^{+05}$ & $1.62 \times 10^{+04}$  & 	$7.19\times 10^{+04}$ & $9.93\times 10^{+03}$
			& $1.78 \times 10^{+05}$ & $1.25 \times 10^{+04}$ \\
			& $30$  & $5.53 \times 10^{+05}$ & $4.65 \times 10^{+04}$ & $2.86\times 10^{+05}$ & $1.98\times 10^{+4}$ & $5.73 \times 10^{+05}$ & $3.18 \times 10^{+04}$ \\
			\bottomrule
		\end{tabular}	}
	\small
	\addtolength{\tabcolsep}{1pt}
	\scalebox{.7}{
		\begin{tabular}{ccccccc}
			\multicolumn{1}{c}{} \\
			\multicolumn{2}{c}{} & \multicolumn{2}{c}{{$^{79}$As}} & \multicolumn{2}{c}{{$^{80}$As}}\\
			\cmidrule(rl){3-4} \cmidrule(rl){5-6}
			$\rho$ & T$_9$  & $\lambda_{pn-QRPA}^{(\beta^{+}+ EC)}$ & $\lambda_{IPM-03}^{(\beta^{+}+EC)}$ & $\lambda_{pn-QRPA}^{(\beta^{+}+ EC)}$ & $\lambda_{IPM-03}^{(\beta^{+}+ EC)}$ \\
			\midrule
			$10^{3}$  & $3$  &	$1.76\times 10^{-11}$ &	$9.73\times 10^{-12}$ &	$7.52\times 10^{-12}$
			& $4.94 \times 10^{-09}$ \\
			& $10$ &	$1.48\times 10^{-03}$ &	$9.94\times 10^{-04}$
			 & $3.72
			\times 10^{-03}$ & $3.96 \times 10^{-03}$ \\
			& $30$  &	$2.08\times 10^{+02}$ &	$2.23\times 10^{+01}$  & $4.11
			\times 10^{+02}$ & $4.01 \times 10^{+01}$ \\
			$10^{7}$  & $3$  &	$9.10\times 10^{-10}$ &	$5.06\times 10^{-10}$
			 & $3.89
			\times 10^{-10}$ & $2.56 \times 10^{-07}$ \\
			& $10$  &	$1.86\times 10^{-03}$ &	$1.25\times 10^{-03}$
			 & $4.68 \times 10^{-03}$ & $4.97 \times 10^{-03}$ \\
			& $30$  &	$2.10 \times 10^{+02}$ & $2.25\times 10^{+01}$
			 & $4.15 \times 10^{+02}$ & $4.04 \times 10^{+01}$ \\
			$10^{11}$  & $3$  &	$4.67\times 10^{+04}$ &	$8.00\times 10^{+03}$
			 & $1.04 \times 10^{+05}$ & $1.06 \times 10^{+04}$ \\
			& $10$ &	$8.63\times 10^{+04}$ &	$8.20\times 10^{+03}$ 
			 & $2.22 \times 10^{+05}$ & $1.07 \times 10^{+04}$ \\
			& $30$  &	$3.99\times 10^{+05}$ &	$1.58\times 10^{+04}$
		 & $7.78\times 10^{+05}$ & $2.38 \times 10^{+04}$ \\
			\bottomrule
		\end{tabular}
	}
\end{table}

\begin{table}[H]
	\centering
	\caption{Same as Table \ref{Tab 05} but for sum of ($\beta^{-}$+$PC$) rates.}
	\label{Tab 05}
	\small
	\addtolength{\tabcolsep}{1pt}
	\scalebox{.7}{
		\begin{tabular}{ccccccccc}
			\toprule
			\multicolumn{1}{c}{} \\
			\multicolumn{2}{c}{} & \multicolumn{2}{c}{{$^{67}$As}} & \multicolumn{2}{c}{{$^{68}$As}} & \multicolumn{2}{c}{{$^{69}$As}}\\
			\cmidrule(rl){3-4} \cmidrule(rl){5-6} \cmidrule(rl){7-8}
			$\rho$ & T$_9$  & $\lambda_{pn-QRPA}^{(\beta^{-}+ PC)}$ & $\lambda_{IPM-03}^{(\beta^{-}+ PC)}$ & $\lambda_{pn-QRPA}^{(\beta^{-}+ PC)}$ & $\lambda_{IPM-03}^{(\beta^{-}+ PC)}$  & $\lambda_{pn-QRPA}^{(\beta^{-}+ PC)}$ & $\lambda_{IPM-03}^{(\beta^{-}+ PC)}$\\
			\midrule
			$10^{3}$  & $3$ &	$1.03\times 10^{-19}$ &	$8.24\times 10^{-17}$  & $4.95 \times 10^{-14}$ & $1.60 \times 10^{-09}$ &	$7.29\times 10^{-15}$ &	$4.18\times 10^{-12}$
			\\
			& $10$  & $5.98\times 10^{-06}$ & $2.88\times 10^{-05}$ & $4.14 \times 10^{-04}$ & $1.62 \times 10^{-04}$  &	$8.36\times 10^{-05}$ &	$1.88\times 10^{-04}$
			\\
			& $30$ &	$6.15\times 10^{+00}$ &	$2.54\times 10^{+00}$ & $3.80 \times 10^{+01}$ & $8.34 \times 10^{+00}$ &	$1.45\times 10^{+01}$ &	$5.46\times 10^{+00}$
			\\
			$10^{7}$  & $3$ &	$1.98\times 10^{-21}$ &	$1.61\times 10^{-18}$
			 & $9.53 \times 10^{-16}$ & $3.12 \times 10^{-11}$ &	$1.41\times 10^{-16}$ &	$8.17\times 10^{-14}$ 
			\\
			& $10$ &	$4.76\times 10^{-06}$ &	$2.30\times 10^{-05}$
			 & $3.30 \times 10^{-04}$ & $1.30 \times 10^{-04}$ & $6.65\times 10^{-05}$ &	$1.50\times 10^{-04}$ 
			\\
			& $30$ &	$6.11\times 10^{+00}$ &	$2.51\times 10^{+00}$
			 & $3.80 \times 10^{+01}$ & $8.34 \times 10^{+00}$ &	$1.43\times 10^{+01}$ &	$5.42\times 10^{+00}$
			\\
			$10^{11}$  & $3$ &	$6.61\times 10^{-60}$ & $5.94\times 10^{-57}$
			 & $3.18 \times 10^{-54}$ & $1.15 \times 10^{-49}$ &	$4.69\times 10^{-55}$ &	$3.01\times 10^{-52}$ 
			\\
			& $10$ &	$5.82\times 10^{-18}$ &	$2.94\times 10^{-17}$
			 & $4.04 \times 10^{-16}$ & $1.67 \times 10^{-16}$ &	$8.15\times 10^{-17}$ &	$1.93\times 10^{-16}$
			\\
			& $30$ &	$8.38\times 10^{-04}$ &	$3.52\times 10^{-04}$
			 & $5.21 \times 10^{-03}$ & $1.19 \times 10^{-03}$ &	$1.97\times 10^{-03}$ &	$7.62\times 10^{-04}$ 
			\\
			\bottomrule
		\end{tabular}
	}
	\small
	\addtolength{\tabcolsep}{1pt}
	\scalebox{.7}{
		\begin{tabular}{ccccccccc}
			\multicolumn{1}{c}{} \\
			\multicolumn{2}{c}{} & \multicolumn{2}{c}{{$^{70}$As}} & \multicolumn{2}{c}{{$^{71}$As}} & \multicolumn{2}{c}{{$^{72}$As}}\\
			\cmidrule(rl){3-4} \cmidrule(rl){5-6} \cmidrule(rl){7-8}
			$\rho$ & T$_9$ & $\lambda_{pn-QRPA}^{(\beta^{-}+ PC)}$ & $\lambda_{IPM-03}^{(\beta^{-}+PC)}$  & $\lambda_{pn-QRPA}^{(\beta^{-}+ PC)}$ & $\lambda_{IPM-03}^{(\beta^{-}+PC)}$  & $\lambda_{pn-QRPA}^{(\beta^{-}+ PC)}$ & $\lambda_{IPM-03}^{(\beta^{-}+ PC)}$ \\
			\midrule
			$10^{3}$ & $3$  & $5.64 \times 10^{-11}$ & $1.88 \times 10^{-07}$  &	$3.88\times 10^{-12}$ &	$1.45\times 10^{-08}$ & $4.83 \times 10^{-08}$ & $9.54 \times 10^{-07}$ \\
			& $10$ & $1.36 \times 10^{-03}$ & $5.36 \times 10^{-04}$  &	$3.33\times 10^{-04}$ &	$1.97\times 10^{-03}$ 
			 & $6.56 \times 10^{-03}$ & $1.25 \times 10^{-03}$ \\
			& $30$   & $4.22 \times 10^{+01}$ & $1.32 \times 10^{+01}$ &	$2.18\times 10^{+01}$ &	$8.26\times 10^{+00}$
			 & $8.73 \times 10^{+01}$ & $1.23 \times 10^{+01}$ \\
			$10^{7}$  & $3$ & $1.08 \times 10^{-12}$ & $3.68 \times 10^{-09}$ &	$7.48\times 10^{-14}$ &	$2.83\times 10^{-10}$ 
			 & $1.49 \times 10^{-09}$ & $1.76 \times 10^{-07}$ \\
			& $10$  & $1.09 \times 10^{-03}$ & $4.29 \times 10^{-04}$ &	$2.65\times 10^{-04}$ &	$1.58\times 10^{-03}$
			 & $5.24 \times 10^{-03}$ & $1.01 \times 10^{-03}$ \\
			& $30$ & $4.18 \times 10^{+01}$ & $1.30 \times 10^{+01}$  &	$2.16\times 10^{+01}$ &	$8.18\times 10^{+00}$
			 & $8.65 \times 10^{+01}$ & $1.21 \times 10^{+01}$ \\
			$10^{11}$  & $3$ & $3.62 \times 10^{-51}$ & $1.35 \times 10^{-47}$ &	$2.49\times 10^{-52}$ &	$1.04\times 10^{-48}$
			 & $1.26 \times 10^{-47}$ & $4.36 \times 10^{-47}$ \\
			& $10$ & $1.33 \times 10^{-15}$ & $5.52 \times 10^{-16}$ &	$3.25\times 10^{-16}$ &	$2.04\times 10^{-15}$
			 & $6.55 \times 10^{-15}$ & $1.80 \times 10^{-15}$  \\
			& $30$ & $5.74 \times 10^{-03}$ & $1.85 \times 10^{-03}$ &	$2.96\times 10^{-03}$ &	$1.15\times 10^{-03}$ 
			 & $1.19 \times 10^{-02}$ & $1.72 \times 10^{-03}$ \\
			\bottomrule
		\end{tabular}
	}
	\small
	\addtolength{\tabcolsep}{1pt}
	\scalebox{.7}{
		\begin{tabular}{ccccccccc}
			\multicolumn{1}{c}{} \\
			\multicolumn{2}{c}{} & \multicolumn{2}{c}{{$^{73}$As}} & \multicolumn{2}{c}{{$^{74}$As}} & \multicolumn{2}{c}{{$^{75}$As}}\\
			\cmidrule(rl){3-4} \cmidrule(rl){5-6} \cmidrule(rl){7-8}
			$\rho$ & T$_9$ & $\lambda_{pn-QRPA}^{(\beta^{-}+ PC)}$ & $\lambda_{IPM-03}^{(\beta^{-}+PC)}$  & $\lambda_{pn-QRPA}^{(\beta^{-}+ PC)}$ & $\lambda_{IPM-03}^{(\beta^{-}+PC)}$  & $\lambda_{pn-QRPA}^{(\beta^{-}+ PC)}$ & $\lambda_{IPM-03}^{(\beta^{-}+ PC)}$ \\
			\midrule
			$10^{3}$ & $3$ &	$7.58\times 10^{-09}$ &	$3.81\times 10^{-06}$
			 & $9.02 \times 10^{-06}$ & $4.69 \times 10^{-05}$  &	$3.68\times 10^{-06}$ &	$5.78\times 10^{-06}$
			\\
			& $10$  & 	$2.51\times 10^{-03}$ & 	$8.36\times 10^{-03}$  & $3.40 \times 10^{-02}$ & $3.32 \times 10^{-03}$ &	$8.72\times 10^{-03}$ &	$1.20\times 10^{-02}$
			\\
			& $30$  & $6.03\times 10^{+01}$ &	$1.03\times 10^{+01}$ & $2.13 \times 10^{+02}$ & $1.46 \times 10^{+01}$ &	$7.01\times 10^{+01}$ &	$1.38\times 10^{+01}$ 
			\\
			$10^{7}$  & $3$ &	$1.51\times 10^{-10}$ &	$7.44\times 10^{-08}$ & $2.92 \times 10^{-07}$ & $2.71 \times 10^{-05}$ &	$1.68\times 10^{-07}$ &	$6.75\times 10^{-07}$
			\\
			& $10$  &	$2.00\times 10^{-03}$ &	$6.68\times 10^{-03}$ 
			 & $2.73 \times 10^{-02}$ & $2.83 \times 10^{-03}$  &	$7.00\times 10^{-03}$ &	$9.70\times 10^{-03}$
			\\
			& $30$ &	$5.97\times 10^{+01}$ &	$1.02\times 10^{+02}$
			 & $2.11 \times 10^{+02}$ & $1.45 \times 10^{+01}$  &	$6.97\times 10^{+01}$ &	$1.37\times 10^{+01}$
			\\
			$10^{11}$  & $3$  &	$5.41\times 10^{-49}$ &	$2.75\times 10^{-46}$
			 & $1.30 \times 10^{-44}$ & $2.57 \times 10^{-46}$  &	$5.34\times 10^{-45}$ &	$3.18\times 10^{-46}$
			\\
			& $10$  &	$2.47\times 10^{-15}$ &	$8.64\times 10^{-15}$
			 & $3.95 \times 10^{-14}$ & $2.16 \times 10^{-14}$ &	$1.00\times 10^{-14}$ &	$1.57\times 10^{-14}$
			\\
			& $30$  & $8.22\times 10^{-03}$ &	$1.43\times 10^{-03}$
			 & $2.92 \times 10^{-02}$ & $2.09 \times 10^{-03}$  &	$9.59\times 10^{-03}$ &	$1.92\times 10^{-03}$ 
			\\
			\bottomrule
		\end{tabular}
	}
	\small
	\addtolength{\tabcolsep}{1pt}
	\scalebox{.7}{
		\begin{tabular}{ccccccccc}
			\multicolumn{1}{c}{} \\
			\multicolumn{2}{c}{} & \multicolumn{2}{c}{{$^{76}$As}} & \multicolumn{2}{c}{{$^{77}$As}} & \multicolumn{2}{c}{{$^{78}$As}}\\
			\cmidrule(rl){3-4} \cmidrule(rl){5-6} \cmidrule(rl){7-8}
			$\rho$ & T$_9$  & $\lambda_{pn-QRPA}^{(\beta^{-}+ PC)}$ & $\lambda_{IPM-03}^{(\beta^{-}+PC)}$  & $\lambda_{pn-QRPA}^{(\beta^{-}+ PC)}$ & $\lambda_{IPM-03}^{(\beta^{-}+PC)}$  & $\lambda_{pn-QRPA}^{(\beta^{-}+ PC)}$ & $\lambda_{IPM-03}^{(\beta^{-}+ PC)}$ \\
			\midrule
			$10^{3}$  & $3$  & $6.06 \times 10^{-05}$ & $2.01 \times 10^{-03}$  & $8.02\times 10^{-05}$ &	$3.85\times 10^{-04}$
			 & $7.82 \times 10^{-03}$ & $2.15 \times 10^{-02}$  \\
			& $10$ & $5.58 \times 10^{-02}$ & $1.26 \times 10^{-02}$  &	$1.91\times 10^{-02}$ &	$3.98\times 10^{-02}$
			 & $1.66 \times 10^{-01}$ & $6.29 \times 10^{-02}$  \\
			& $30$ & $1.99 \times 10^{+02}$ & $1.94 \times 10^{+01}$  &	$1.01\times 10^{+02}$ &	$1.79\times 10^{+01}$ 
			 & $2.74 \times 10^{+02}$ & $2.55 \times 10^{+01}$ \\
			$10^{7}$  & $3$ & $1.38 \times 10^{-05}$ & $6.35 \times 10^{+02}$  & $2.29\times 10^{-05}$ & $2.00\times 10^{-04}$ 
			 & $5.08 \times 10^{-03}$ & $1.92 \times 10^{-02}$ \\
			& $10$  & $4.54 \times 10^{-02}$ & $1.15 \times 10^{-02}$  & $1.54\times 10^{-02}$ &	$3.42\times 10^{-02}$
			 & $1.40 \times 10^{-01}$ & $5.98 \times 10^{-02}$ \\
			& $30$  & $1.97 \times 10^{+02}$ & $1.92 \times 10^{+01}$  &	$1.00\times 10^{+02}$ &	$1.78\times 10^{+01}$
			 & $2.72 \times 10^{+02}$ & $2.53 \times 10^{+01}$ \\
			$10^{11}$  & $3$ & $2.00 \times 10^{-42}$ & $1.37 \times 10^{-45}$  &	$1.76\times 10^{-42}$ &	$1.91\times 10^{-45}$ 
			 & $1.91 \times 10^{-39}$ & $3.95 \times 10^{-45}$ \\
			& $10$  & $1.15 \times 10^{-13}$ & $3.63 \times 10^{-13}$ &	$3.45\times 10^{-14}$ &	$2.13\times 10^{-13}$ 
			 & $1.12 \times 10^{-12}$ & $4.84 \times 10^{-12}$ \\
			& $30$  & $2.72 \times 10^{-02}$ & $3.05 \times 10^{-03}$  &	$1.39\times 10^{-02}$ &	$2.52\times 10^{-03}$ 
			 & $3.77 \times 10^{-02}$ & $4.82 \times 10^{-03}$ \\
			\bottomrule
		\end{tabular}
	}
	\addtolength{\tabcolsep}{1pt}
	\scalebox{.7}{
		\begin{tabular}{ccccccc}
			\multicolumn{1}{c}{} \\
			\multicolumn{2}{c}{} & \multicolumn{2}{c}{{$^{79}$As}} & \multicolumn{2}{c}{{$^{80}$As}}\\
			\cmidrule(rl){3-4} \cmidrule(rl){5-6}
			$\rho$ & T$_9$ & $\lambda_{pn-QRPA}^{(\beta^{-}+ PC)}$ & $\lambda_{IPM-03}^{(\beta^{-}+ PC)}$  & $\lambda_{pn-QRPA}^{(\beta^{-}+ PC)}$ & $\lambda_{IPM-03}^{(\beta^{-}+ PC)}$ \\
			\midrule
			$10^{3}$  & $3$  &	$3.68\times 10^{-06}$ &	$1.17\times 10^{-02}$
			 & $1.19 \times 10^{-01}$ & $1.14 \times 10^{-01}$ \\
			& $10$  &	$8.72\times 10^{-03}$ &	$1.73\times 10^{-01}$ 
			 & $6.75 \times 10^{-01}$ & $2.76 \times 10^{-01}$ \\
			& $30$ &	$7.01\times 10^{+01}$ &	$2.34\times 10^{+01}$ 
			 & $4.88 \times 10^{+02}$ & $3.54 \times 10^{+01}$ \\
			$10^{7}$ & $3$  &	$1.68\times 10^{-07}$ &	$9.27\times 10^{-03}$ 
			 & $1.03 \times 10^{-01}$ & $1.05 \times 10^{-01}$ \\
			& $10$ &	$7.00\times 10^{-03}$ &	$1.62\times 10^{-01}$ 
			 & $6.07 \times 10^{-01}$ & $2.67 \times 10^{-01}$ \\
			& $30$  &	$6.97\times 10^{+01}$ &	$2.32\times 10^{+01}$
			 & $4.84 \times 10^{+02}$ & $3.52 \times 10^{+01}$ \\
			$10^{11}$  & $3$  &	$5.34\times 10^{-45}$ &	 $5.87\times 10^{-45}$
			 & $1.95 \times 10^{-36}$ & $9.91 \times 10^{-45}$ \\
			& $10$  &	$1.00\times 10^{-13}$ &	$4.12\times 10^{-12}$
			 & $1.42 \times 10^{-11}$ & $5.16 \times 10^{-11}$ \\
			& $30$  &	$9.59\times 10^{-03}$ &	$3.41\times 10^{-03}$  & $6.78 \times 10^{-02}$ & $9.65 \times 10^{-03}$ \\
			\bottomrule
		\end{tabular}
	}
\end{table}

\section{Conclusion}
The RMF model was used to investigate selected nuclear structure properties of As nuclei including PECs and deformation parameter. The calculations were performed using the DD-ME2 interaction. Our model predicted oblate shapes for $^{67-75}$As and prolate configurations for $^{76-80}$As. The $\beta_{2}$ values were computed using the RMF model and later used as an input parameter in the  pn-QRPA model to perform self-consistent calculations of the $\beta$-decay properties of  As isotopes in the mass region $67 \leq A \leq 80$. The calculated  GT distributions were found to be in reasonable agreement with the measured data and computed half-lives were within a factor 10 of the measured ones. The calculated log $\textit{ft}$ values were compared with previous calculations and measurements   (wherever available) and a decent agreement was reported with the measured data.

The stellar weak interaction rates for As isotopes were computed in a total microscopic fashion without assuming the Brink-Axel hypothesis in  calculation of excited states GT distributions. The pn-QRPA calculated weak rates were compared with the previous shell model rates. The reported weak rates are bigger than the previous calculations by as much as factor of 33. The difference is attributed to the usage of Brink-Axel hypothesis and incorporation of quenching factor in shell model calculation. The reported stellar rates could prove useful for $r$-process nucleosynthesis calculations and simulations of late time stellar evolution. 
\section*{Acknowledgements}
A. Kabir would like to acknowledge the useful discussion with Prof. Peter Ring.\\ J.-U. Nabi  and S. A. Rida would like to acknowledge the support of the Higher Education Commission Pakistan through project number 20-15394/NRPU/R\&D/HEC/2021. \
\section*{Declaration of competing interest}
The authors declare that they have no known competing financial interests or personal relationships that could have appeared to
influence the work reported in this paper.

\end{document}